\documentstyle[titlepage,12pt]{article}

\begin{document}

\renewcommand{\thesection}{\Roman{section}.} \baselineskip=24pt plus 1pt
minus 1pt

\begin{titlepage}

\vspace*{0.5cm}

\begin{center}
\LARGE\bf Quantum Recurrences in Driven Power-law Potentials 
\\[1.5cm]
\normalsize\bf Shahid Iqbal$^1$, Qurat-ul-Ann$^1$ and Farhan Saif$^{1,2*}$
\end{center}

\vspace{7pt}
\begin{description}
%\normalsize\it
\item [$^1$] Department of Electronics, Quaid-i-Azam University, Islamabad 45320, Pakistan.
\item [$^2$] Department of Physics and Astronomical Sciences , The University of Arizona, Tucson 85721, Arizona, USA.
\item [$^*$] saif@physics.arizona.edu,saif@fulbrightweb.org
\end{description}

\vspace{0.3cm}

\begin{center}
\normalsize 
The recurrence phenomena of an initially well localized wave packet
are studied in periodically driven power-law potentials. For our
general study we divide the potentials in two kinds, namely tightly binding
and loosely binding potentials. In the presence of an external
periodically modulating force, these potentials may exhibit 
classical and quantum chaos. The dynamics of a quantum wave packet
in the modulated potentials displays recurrences at various time
scales. We develop general analytical relations for these times and
discuss their parametric dependence.
\end{center}

\vspace{0.2cm} \noindent PACS numbers: 05.45.-a, 03.65.-w, 45.80.+r, 47.52.+j

%\vspace{0.2cm} \noindent Keywords: dynamical systems, recurrences,quantum revivals, quantum chaos

%\vspace{0.6cm} \noindent Fax: +92 51 9210256

%\vspace{0.4cm}

%\noindent Tel.: +92 51 2821617

%\vspace{0.4cm}

%\noindent E.mail: farhan@qau.edu.pk, saif@physik.uni-ulm.de

\end{titlepage}

\newpage

\section{\label{sec:level1}Introduction}

In one degree of freedom systems, wave packet dynamics manifests
quantum recurrences~\cite{one,three,four,five,six,seven,eight,nine,niner}.
An initially well-localized wave packet in a bounded system 
follows classical evolution
in its short time dynamics and displays reconstruction after a
classical period. However, after many classical
periods phase difference between constituent wavelets develops which
leads to destructive interference and results a collapse when
the phase difference is maximum. Later, constructive interference
dominates and supports quantum revivals and fractional revivals.
In this paper we study a general class of
one degree of freedom bounded systems defined by power-law potentials 
and show that the quantum recurrence phenomena occurs in the presence of 
external periodic forces.

The quantum recurrences have been studied in a variety of quantum
mechanical systems, such as, in the Jaynes-Cummings model of the
quantum electrodynamics~\cite{ten,elev}, in the micromaser
cavity~\cite{twel}, in the Rydberg wave packets~\cite{thir}, and in
the multi-atomic molecules~\cite{fort}. Quantum recurrences in periodically 
driven quantum chaotic systems~\cite{hogg,fish,fift,twen} 
and in general higher 
dimensional systems are proved to be generic~\cite{sixt,seve}.
The phenomena is vital to extend the horizons of nanotechnology and to
develop newer experimental techniques to 
study surface structures with atomic size resolution~\cite{RTM}.
 
In this paper, we introduce a general class of
potentials as power-law potentials~\cite{suk,nieto,liboff,eigh} and calculate the
quantum recurrence times in the presence of external periodic
force. We develop analytical relations for recurrence times, namely
the classical period and the quantum revival time, and explain their
parametric dependence. Our analytical treatment leads us to explain
the interdependence of these times in the loosely binding and the
tightly binding power law potentials as well.

The paper is organized as follows. In Sec.~\ref{sec:cop}, we
classify the general power law potentials and write
the general Hamiltonian and corresponding quantized energy for
the system. In Sec.~\ref{sec:qrplp}, we calculate the general
relations for the classical period and the quantum revival time
in the power-law potentials. In Sec.~\ref{ple}, we explain the
dependence of time scales on nonlinearity-measure parameter. Later,
we extend the discussion to tightly binding potentials and loosely
binding potentials in Secs.~\ref{tbp} and \ref{lbp}, respectively.
%In Sec.~ \ref{sec:mgc} we study the modulated gravitational cavity
%as an example of loosely binding potentials.

\section{Classification of potentials}
\label{sec:cop}

On the basis of their energy spectrum, we classify potentials as of
two kinds: \textit{i)} Tightly binding potentials,
for which level spacing between the adjacent energy levels increases
as the quantum number increases. \textit{ii)} Loosely binding potentials, 
for which the
level spacing between adjacent levels reduces as the principle
quantum number increases.
We may express the two kinds of the
potentials mathematically by a general power law expression, viz.,
\begin{equation}
\tilde{V}_{(k)}(X)=\tilde{V}_{0}\left| \frac{X}{a}\right| ^{k}, 
\label{vkx}
\end{equation}
where,  
${\tilde V}_{0}$ and $a$ are constants
with the dimensions of energy and length, respectively.
Moreover, $k$ defines the power-law exponent. 
For the exponent $k>2$, we find tightly binding potentials 
and $k<2$ expresses loosely binding potentials.

In the presence of an external time periodic force, the dynamics in
power-law potentials is governed by the general Hamiltonian given as
\begin{equation}
\tilde{H}(X,P,\tau )=\tilde{H}_{0}(X,P)+\lambda \tilde{V}(X)\sin \Omega \tau
,  \label{eq1}
\end{equation}%
where, $\tilde{H}_{0}(X,P)=\frac{P^{2}}{2m}+\tilde{V}_{0}\left| \frac{X}{a}%
\right| ^{k}$, describes the unmodulated system. In the second term
$\lambda $ expresses the dimensionless external
modulation strength.

With the help of WKB approximation we express the energy 
quantization in the unmodulated system~\cite{suk, nieto, liboff, eigh} as
\begin{equation}
\tilde{E}_{n}^{(k)}=\left[ \left(n+\frac{\gamma }{4}\right)\frac{\hbar \pi }{2a\sqrt{2m}%
}\tilde{V}_{0}^{1/k}\frac{\Gamma (1/k+3/2)}{\Gamma (1/k+1)\Gamma (3/2)}%
\right] ^{2k/(k+2)},  \label{plen}
\end{equation}
here, $n$ is the corresponding principle quantum number. Moreover, $\gamma $
defines the Maslov index, which takes integer values depending on the
boundary conditions at the turning points. For example, in the case of
harmonic oscillator we have $\gamma=2$ and in case
of one dimensional box we find $\gamma=4$.

Equation~(\ref{plen}) leads us to conclude that the energy difference
between adjacent
levels, $\Delta E_{n}=E_{n}-E_{n-1}\propto (n+\frac{\gamma }{4})^{\frac{%
k-2}{k+2}}$, increases for $k>2$ as the quantum number $n$
increases. Therefore, Eq.~(\ref{vkx}) defines tightly binding
potentials for $k>2$. In contrast, for $k<2$ the energy
difference between adjacent levels, $\Delta E_{n}=E_{n}-E_{n-1}\propto (n+%
\frac{\gamma }{4})^{\frac{k-2}{k+2}}$, decreases with the increase of the
principle quantum number $n$. For the reason, Eq.~(\ref{vkx}) defines loosely binding
potentials for $k<2$.

In order to simplify our calculations we introduce the dimensionless
position coordinate, $x=\frac{X}{a}$, and dimensionless momentum coordinate,
$p=\frac{P}{\sqrt{m\hbar \Omega }}$. The commutation relation between $x$
and $p$ leads us to find out an effective Planck's constant, $k^{\hspace{%
-2.1mm}-}$, such that,
\begin{equation}
\lbrack x,p]={\frac{1}{a\sqrt{m\Omega \hbar }}}[X,P]=ik^{\hspace{%
-2.1mm}-},
\end{equation}%
where, $k^{\hspace{-2.1mm}-}=\frac{1}{a}\sqrt{\frac{\hbar }{m\Omega }}$.
The scaled position and momentum coordinates help us to scale the general 
Hamiltonian given in Eq.~(\ref{eq1}), and we find
\begin{equation}
H(x,p,t)=\frac{p^{2}}{2}+V_{0}\left| x\right| ^{k}+\lambda V(x)\sin t\equiv
H_{0}(x,p)+\lambda V(x)\sin t,  \label{eq4}
\end{equation}%
where, $H(x,p,t)$ = $\tilde{H}(X,P,\tau )/\hbar \Omega $, and $%
t=\Omega \tau $. Moreover, the amplitude 
$V_{0}={\tilde{V}_{0}}/{\hbar \Omega }$
and the spatial component of external modulation 
$V(x)={\tilde{V}(X)}/{\hbar \Omega }$.
In the dimensionless form the quantized energy, given in Eq.~(\ref{plen}),
becomes
\begin{equation}
E_{n}^{(k)}=\left[\left(n+\frac{\gamma }{4}\right)
\frac{k^{\hspace{-2.1mm}-}\pi }{2%
\sqrt{2}}V_{0}^{1/k}\frac{\Gamma (1/k+3/2)}{\Gamma (1/k+1)\Gamma (3/2)}%
\right] ^{2k/(k+2)},  \label{eq5}
\end{equation}%
where, $E_{n}^{(k)}=\tilde{E}_{n}^{(k)}/\hbar \Omega $, is the scaled energy.

%%%%%%%%%%%%%%%%%%%%%%%%%%%%%%%%%%%%%%%%%%%%%%%%%%%%%%%%%%%%%%%%%%%%%%%%%%%

\section{Quantum Recurrences in Power law potentials}

\label{sec:qrplp}

In the periodically driven power-law potentials energy is no longer a
constant of the motion. For the reason we solve the time dependent
Schr\"{o}dinger equation by using secular perturbation approximation
as suggested by Max Born~\cite{nint}. This leads us to find the solution 
by averaging over rapidly changing variable and we, therefore, 
find out a partial solution of the
periodically driven system. The procedure provides us the quasi eigen energy
states and the quasi eigen energy values for the nonlinear
resonances of the system. %Later, we may expand the initial wave packet projected
%in the driven system over the quasi energy eigen states~\cite{twen}.

In order to study the $N$th  
quantum nonlinear resonances of the system~\cite%
{twen1,twen2}, we write the solution of 
the Schr\"{o}dinger equation, corresponding to
the Hamiltonian in Eq.~(\ref{eq4}),  
in the form of an anstaz, viz.,
\begin{equation}
|\psi (t)\rangle =\sum_{n}C_{n}(t)|n\rangle \exp \left\{ -i\left[ E_{\bar{n}%
}^{(k)}+(n-\bar{n})\frac{k^{\hspace{-2.1mm}-}}{N}\right] \frac{t}{k^{\hspace{%
-2.1mm}-}}\right\} .  \label{psit}
\end{equation}%
Here, $E_{\bar{n}}^{(k)}$ is the mean energy of the wave packet in the Nth
resonance.

On substituting Eq.~(\ref{psit}) in the time dependent Schr\"{o}dinger
equation, we find that the probability amplitude for any nth state, $C_{n},$
changes following the equation,
\begin{equation}
ik^{\hspace{-2.1mm}-}{\dot{C}}_{n}=\left[ E_{n}^{(k)}-E_{\bar{n}}^{(k)}-(n-%
\bar{n})\frac{k^{\hspace{-2.1mm}-}}{N}\right] C_{n}(t)+\frac{\lambda V}{2i}%
(C_{n+N}-C_{n-N}).  \label{cdot}
\end{equation}%
In order to obtain Eq.~(\ref{cdot}), we average over the fast
oscillating terms and keep only the resonant ones. For larger values
of $n$, we consider the corresponding off-diagonal matrix elements
approximately the same, so that, $V_{n,n+N}\approx V_{n,n-N}=V$.

We take the initial excitation, narrowly peaked around the mean
value, $\bar{n}$. For the reason, we take slow variations in the
energy, $E_{n}^{(k)}$, around the $\bar{n}$ in a nonlinear
resonance, and expand it up to second order in Taylor's expansion.
Thus, the equation of motion for the probability amplitude, $C_{n}$,
is
\begin{equation}
ik^{\hspace{-2.1mm}-}{{\dot{C}}_{n}}=k^{\hspace{-2.1mm}-}(n-\bar{n})(\omega -%
\frac{1}{N})C_{n}(t)+\frac{1}{2}k^{\hspace{-2.1mm}-}(n-\bar{n})^{2}\zeta
C_{n}(t)+\frac{\lambda V}{2i}(C_{n+N}-C_{n-N}).  \label{prob}
\end{equation}%
Here, the frequency of the system $\omega $ is defined by
\begin{equation}
\omega =\frac{\partial E_{n}^{(k)}}{k^{\hspace{-2.1mm}-}\partial n}|_{n=\bar{%
n}}=\frac{1}{k^{\hspace{-2.1mm}-}}\frac{2k}{k+2}(\bar{n}+\frac{\gamma }{4}%
)^{-1}E_{\bar{n}}^{(k)},  \label{omeg}
\end{equation}%
and the nonlinearity, $\zeta $, in the time independent energy
spectrum becomes
\begin{equation}
\zeta =\frac{\partial ^{2}E_{n}^{(k)}}{{k^{\hspace{-2.1mm}-}}^{2}\partial
n^{2}}|_{n=\bar{n}}=\frac{1}{{k^{\hspace{-2.1mm}-}}^{2}}\frac{2k(k-2)}{%
(k+2)^{2}}(\bar{n}+\frac{\gamma }{4})^{-2}E_{\bar{n}}^{(k)}.  \label{zeta}
\end{equation}

We introduce the Fourier representation for $ C_{n}$ as,
\begin{equation}
C_{n}=\frac{1}{2N\pi }\int_{0}^{2N\pi }g(\theta, t)e^{-i(n-\bar{n})\theta
/N}d\theta ,  \label{four}
\end{equation}%
which helps us to express Eq.~(\ref{prob}) as the Schr\"odinger
equation for $g(\theta,t )$, such that
$ik^{\hspace{-2.1mm}-}{\dot{g}}(\theta ,t)=H(\theta )g(\theta ,t)$~\cite{twen1,twen2}.
The Hamiltonian $H(\theta )$ is, therefore, given as,
\begin{equation}
H(\theta )=-\frac{N^{2}{k^{\hspace{-2.1mm}-}}^{2}\zeta }{2}\frac{\partial
^{2}}{\partial \theta ^{2}}-iNk^{\hspace{-2.1mm}-}(\omega -\frac{1}{N})\frac{%
\partial }{\partial \theta }-\lambda V\sin \theta .  \label{thet}
\end{equation}%
In order to obtain this equation, we consider the function $g(\theta
,t)$ as $2N\pi$ periodic, in $\theta$ coordinate.

Due to the time-independent behavior of the Hamiltonian, we can write the
time evolution of $g(\theta ,t)$, as $g(\theta ,t)=\tilde{g}(\theta )\exp\{
-i\varepsilon t/k^{\hspace{-2.1mm}-}\}$. As a consequence, the time
dependent Schr\"{o}dinger equation for $g(\theta ,t)$ , reduces to the
standard Mathieu equation,
\begin{equation}
\left[ \frac{\partial ^{2}}{\partial z^{2}}+a_{\nu }-2q\cos 2z\right] \chi
(z)=0.  \label{math}
\end{equation}%
Here, we express $\tilde{g}=\chi (z)\exp\left(-2i(\omega -
\frac{1}{N}) z/N{k^{\hspace{-2.1mm}-}}\zeta \right)$, where $\theta
=2z+\pi /2$. In Eq.~(\ref{math}), the Mathieu characteristic
parameter $a_{\nu }$, and $q$ are
\begin{equation}
a_{\nu }=\frac{8}{N^{2}{k^{\hspace{-2.1mm}-}}^{2}\zeta }\left[ \frac{(\omega
-\frac{1}{N})^{2}}{2\zeta }+\varepsilon \right] ,  \label{anew}
\end{equation}%
and
\begin{equation}
q=\frac{4\lambda V}{N^{2}{k^{\hspace{-2.1mm}-}}^{2}\zeta },  \label{q}
\end{equation}
respectively.
Hence, the quasi eigen energy of the system is obtained from
Eq.~(\ref{anew}) as,
\begin{equation}
\varepsilon _{\nu }=\frac{N^{2}{k^{\hspace{-2.1mm}-}}^{2}\zeta }{8}a_{\nu }-%
\frac{(\omega -\frac{1}{N})^{2}}{2\zeta },  \label{ener}
\end{equation}
where, the index $\nu$ is $\nu={2(n-\bar
n)}/{N^2}+ {2(\omega-1/N)}/{N}k^{\hspace{-2.1mm}-}\zeta$.

The time scales, $T^{(j)},$ at which recurrences of an initially
well localized wave packet occur in the driven system, depend upon
the quasi eigen energy such that, $T^{(j)}=\frac{2\pi }{\omega
^{(j)}}$, where $\omega
^{(j)}=(j!k^{\hspace{-2.1mm}-})^{-1}\partial^{j}\varepsilon_{n}/\partial
n^j$~\cite{sixt,seve}. Therefore, the classical period, $T^{(1)}$ =$%
T_{\lambda }^{(cl)},$ and the quantum revival time,
$T^{(2)}=T_{\lambda }^{(Q)},$ for the driven power-law potentials 
take the shape
\begin{equation}
T_{\lambda }^{(cl)}=\{1-M_{0}^{(cl)}\}T_{0}^{(cl)}\Delta ,  \label{Tcla}
\end{equation}%
\ and%
\begin{equation}
T_{\lambda }^{(Q)}=\{1-M_{0}^{(Q)}\}T_{0}^{(Q)}.  \label{Tqua}
\end{equation}
Here, $T_{0}^{(cl)}={2\pi }/{\omega }$ and
$T_{0}^{(Q)}={2\pi }/\left( \frac{1}{2!}k^{\hspace{-2.1mm}-}\zeta \right)$
are the classical period and quantum revival time in the
\textit{absence} of external modulation. Moreover, $\Delta
=(1-\omega _{N}/\omega )^{-1}$ where $\omega _{N}=1/N.$

The time modification factors $M_{0}^{(cl)}$ and $M_{0}^{(Q)}$ are given as,
\begin{equation}
M_{0}^{(cl)}=-\left\{ \frac{1}{2}\left( \frac{\lambda V\zeta \Delta ^{2}}{%
\omega ^{2}}\right) ^{2}\frac{1}{(1-\mu ^{2})^{2}}\right\} ,  \label{Mcla}
\end{equation}%
and
\begin{equation}
M_{0}^{(Q)}=\left\{ \frac{1}{2}\left( \frac{\lambda V\zeta \Delta ^{2}}{%
\omega ^{2}}\right) ^{2}\frac{3+\mu ^{2}}{(1-\mu ^{2})^{3}}\right\} ,
\label{Mqua}
\end{equation}%
where $\mu ={Nk^{\hspace{-2.1mm}-}\zeta \Delta }/{2\omega }$. Here, $%
\omega =\frac{\triangle E_{n}^{(k)}}{k^{\hspace{-2.1mm}-}\triangle n}|_{n=%
\bar{n}}$ gives us the information about the energy difference
between two adjacent levels and, $\zeta $ determines the
nonlinearity associated with the energy spectrum corresponding to 
the {\it unmodulated} system.

\section{Power Law Exponent and Quantum Recurrence Times}
\label{ple}

In section~\ref{sec:qrplp}, we calculated the classical period and 
the quantum revival time for the power-law potentials in the presence
of external fields. As discussed 
earlier in section~\ref{sec:cop} we may consider two kinds of
power-law potentials, namely, tightly binding potentials and
loosely binding potentials. The two kinds of
potentials are defined in terms of power law exponent. The typical
value of the exponent reflects itself in the frequency $\omega$, and
in the nonlinearity, $\zeta$, present in the time independent
systems, and expressed in Eqs.~(\ref{omeg}) and (\ref{zeta}), respectively.

We note that for the special case of the quadratic potential the
power-law exponent $k$ takes the value as, $k=2$. 
Equation~(\ref{plen}) provides a non-trivial result and we note that 
in this case the
energy spectrum of the system is linear. Hence,
the spacing between the adjacent levels is fixed to a
constant.
%
%Thus, the Eq.~(\ref{omeg}) explains that the frequency of a particle
%undergoing harmonic oscillations in the quadratic potential becomes
%independent of mean quantum number $\bar n$ and a constant.
%Furthermore, as appears from Eq.~(\ref{zeta}), in the particular
%case the nonlinearity $\zeta$ reduces to zero. Thus the classical
%period, $T_0^{(cl)}$, becomes a constant and the quantum revival
%tim $T_0^{(Q)}$, appears at infinity.
%
Interestingly, a non-zero nonlinearity enters in the energy spectrum
as we go beyond the quadratic potential, and the exponent takes a
value other than $k=2$. Therefore, for the sake of clearity, we define
a nonlinearity-measure parameter $\rho$,
and redefine the power law exponent, $k$, as $k=2+\rho$. It is
obvious that for the quadratic potential, we find $\rho=0$.
Hence, the parameter $\rho$ defines the extent of nonlinearity in a system. 

From Eqs.~(\ref{omeg}) and (\ref{zeta}) we note that in addition to 
the power-law exponent $k$, frequency $\omega$ and
nonlinearity $\zeta$, have important dependence on mean quantum
number $\bar{n}$, and the effective Planck's constant
$k^{\hspace{-2.1mm}-}$. Hence, the general relations for the
frequency, $\omega $, and the nonlinearity, $\zeta$ become
\begin{equation}
\omega =\frac{1}{k^{\hspace{-2.1mm}-}}\frac{2(2+\rho )}{(4+\rho )}\left(\bar{n%
}+\frac{\gamma }{4}\right)^{-1}E_{\bar{n}}^{(\rho )},  \label{omeg1}
\end{equation}%
and%
\begin{equation}
\zeta =\frac{1}{2{k^{\hspace{-2.1mm}-}}^{2}}\frac{\rho (2+\rho )}{%
(4+\rho )^{2}}\left(\bar{n}+\frac{\gamma }{4}\right)^{-2}E_{\bar{n}}^{(\rho )}.
\label{zet1}
\end{equation}

In terms of the nonlinearity measure $\rho$, the Eqs.~(\ref{omeg1})
and~(\ref{zet1}) lead us to calculate the recurrence times,
$T_{\lambda}^{(cl)}$ and $T_{\lambda}^{(Q)}$, given in
Eqs.~(\ref{Tcla}) and~(\ref{Tqua}) in the {\it presence} of external
modulation. We express these times as
\begin{equation} T_{\lambda
}^{(cl)}=\left\{ 1+\frac{1}{2}\left( \frac{\lambda V\rho \Delta
^{2}}{2(2+\rho )E_{\bar{n}}^{(\rho )}}\right) ^{2}\frac{1}{(1-\mu
^{2})^{2}}\right\} T_{0}^{(cl)}\Delta , \label{Tlamcl}
\end{equation}%
and
\begin{equation}
T_{\lambda }^{(Q)}=\left\{ 1-\frac{1}{2}\left( \frac{\lambda V\rho
\Delta
^{2}}{2(2+\rho )E_{\bar{n}}^{(\rho )}}\right) ^{2}\frac{3+\mu ^{2}}{%
(1-\mu ^{2})^{3}}\right\} T_{0}^{(Q)}.  \label{Tlamqu}
\end{equation}%
Here, we have
\begin{equation}
T_{0}^{(cl)}=\frac{\pi k^{\hspace{-2.1mm}-}}{E_{\bar{n}}^{(\rho)}}\frac{%
(4+\rho )}{(2+\rho )}\left(\bar{n}+\frac{\gamma }{4}\right), \label{Tnotcl}
\end{equation}%
and
\begin{equation}
T_{0}^{(Q)}=\frac{2\pi k^{\hspace{-2.1mm}-}}{E_{\bar{n}}^{(\rho)}}\frac{%
(4+\rho )^{2}}{\rho (2+\rho )}\left(\bar{n}+\frac{\gamma }{4}\right)^{2},
\label{Tnotqu}
\end{equation}%
respectively, as the classical period and the quantum revival time in the {\it
absence} of external modulation. Moreover, the parameter $\mu $ is
defined as a function of $\rho$, as
\begin{equation}
\mu =\frac{N\rho \Delta }{2(4+\rho )(\bar{n}+\frac{\gamma }{4})}.
\label{munew}
\end{equation}

{\it Quadratic Potential:} As discussed above, for the quadratic potential 
we find the nonlinearity-measure parameter $\rho=0$, and threfore the energy
spectrum is linear. From
Eqs.~(\ref{Tlamcl}) and~(\ref {Tlamqu}), we note that in this
special case, the time scales in the {\it presence} of the external
modulation and its {\it absence} are related
as,
\begin{eqnarray} 
T_{\lambda }^{(cl)}&=&T_{0}^{(cl)}\Delta,\nonumber\\ 
T_{\lambda}^{(Q)}&=&T_{0}^{(Q)}.
\end{eqnarray}

Interestingly, the Eq.~(\ref{Tnotqu}) leads us to conclude that 
in the quadratic potential case, $T_0^{(Q)}=\infty$, therefore, 
quantum revival in the absence and in the presence of the external modulation 
occurs after an infinite time, whereas the
classical period is constant and independent of $\bar n$. Hence, the
quantum evolution of a particle undergoing harmonic motion in the
quadratic potential, mimics the classical evolution and shows a
recurrence after a classical period only.

The nonlinearity measure $\rho$ may take positive or negative
values. In our later discussion, we consider two possible cases on
$\rho$: First, when $\rho$ is positive which corresponds to tightly
binding potentials; Second, when $\rho$ is negative which expresses
loosely binding potentials. Thus, the quadratic potential provides a boundray 
between the two kind of potentials.

\section{Tightly Binding Potentials}
\label{tbp}

In case of tightly binding potentials, the nonlinearity measure
$\rho,$ is positive. As the value of $\rho$ increases slightly
beyond zero, the energy spectrum exhibits a weak nonlinearity, which
we may express as $\rho =+\epsilon$. From Eq.~(\ref{Tlamcl}), we
note that for the weak nonlinearity case, the classical period in the
presence of the external modulation, $T^{(cl)}_{\lambda}$, becomes
larger than its value in the absence of the nonlinearity. However,
Eq.~(\ref{Tlamqu}) explains that the corresponding value of quantum
revival time, $T^{(Q)}_{\lambda}$, starts reducing from its value
present in the linear case.
Hence, 
from Eqs.~(\ref{Tlamcl}) and (\ref{Tlamqu}) we note that, as the
value of $\rho$ increases, the value of $T^{(cl)}_{\lambda}$ goes on
increasing. However, $T^{(Q)}_{\lambda}$ shows a decrease, so far as
$\mu^2<1$. 

For the smaller values of $\rho$, we find that the classical period
and the quantum revival time in the presence and in the absence of external driving force 
are related as 
\begin{equation}
\frac{3T_{\lambda}^{(cl)}}{4T_0^{(cl)}}+\frac{T_{\lambda}^{(Q)}}{4T_0^{(Q)}}=1.
\end{equation} 
This explains that 
the decrease in quantum
revival time is three times more as compared with the increase in
the classical period in the presence of external
modulation~\cite{sixt}.

At asymptotically large values of $\rho$, we note that the recurrence times
%$T^{(cl)}_{\lambda}$ and $T^{(Q)}_{\lambda}$,
in the presence and in the absence of the external modulation
becomes independent of $\rho$. In the absence of the
modulation the classical period
%depends on the mean quantum number
%$\bar n$ and the effective Planck's constant $k^{\hspace{
%-2.1mm}-}$. However, at present, the revival time depends only on
%$k^{\hspace{ -2.1mm}-}$, as we find in one dimensional
%box~\cite{Marzoli 1998}.
%
%For a fixed value of $\rho$ and the effective Planck's constant,
%$k^{\hspace{ -2.1mm}-}$, we note that
$T_0^{(cl)}$, as given in~(\ref{Tnotcl}), is inversely proportional
to the mean quantum number $\bar{n}$, such that,
$T_0^{(cl)}\propto\left(\bar n +
\gamma/4\right)^{-\frac{\rho}{4+\rho}}$. However, $T_0^{(Q)}$,
displays a different dependence on the mean quantum number,
$\bar{n}$. From Eq.~(\ref{Tnotqu}), we find that the quantum revival
time is directly proportional to ${\bar n}$, such that,
$T_0^{(Q)}\propto \left(\bar n +
\gamma/4\right)^{+\frac{4}{4+\rho}}$. As $\rho$ becomes infinite we
note that $T_0^{(Q)}$ becomes constant for a constant value of
$k^{\hspace{ -2.1mm}-}$ and independent of $\bar{n}$, as we find in
one dimensional box~\cite{Marzoli 1998}.

It is interesting to note that, from Eq.~(\ref{munew}), the
parameter $\mu$ and the mean quantum number $\bar n$ are inversely
proportional to each other. Hence, in tightly binding potentials 
for a smaller (larger) value of
$\bar n$, we find larger (smaller) value for $\mu$.
Equation~(\ref{Tlamqu}), leads to the conclusion
that the quantum revival time $T^{(Q)}_{\lambda}$, increases for the
smaller $\bar n$, and reduces for the larger $\bar n$ in the 
presence of external modulation as compared
with the quantum revival time for the time independent system.
However, the classical period, $T_{\lambda}^{(cl)}$, demonstrate always an 
increase with the modulation strength, as we find from Eq.~(\ref{Tlamcl}). 

For fixed values of $\rho$, and $\bar{n}$, we note that the recurrence
times change as a function of effective Planck's constant,
$k^{\hspace{ -2.1mm}-}$. The Eqs.~(\ref{Tnotcl}) and
(\ref{Tnotqu}), indicate that the times $ T^{(cl)}_{0}$, and
$T^{(Q)}_{0}$, are inversely proportional to
${k^{\hspace{-2.1mm}-}}^{-\rho/(4+\rho)}$. Hence, in potentials with
very weak non-linearity, we find that the value of the Planck's
constant does not contribute appreciably 
to recurrence times. However, for large
value of $\rho$ the recurrence times in the absence of modulation
depend inversely on $k^{\hspace{-2.1mm}-}$.

\section{Loosely binding potentials}
\label{lbp}

For loosely binding potentials the nonlinearity-measure parameter, $\rho$, is
negative, therefore, we may express the exponent $k$ as $k=2-\rho$. 
As discussed earlier, in this case 
the energy spacing between adjacent levels
reduces as the principle quantum number $n$ increases. 
We express a weak nonlinearity in the energy spectrum
by assigning a small value to the nonlinearity-measure parameter.
%, $\rho$, such that $\rho =-\epsilon$, and $\epsilon$ expresses a small number.
The classical period
$T^{(cl)}_{\lambda}$, and the quantum revival time
$T^{(Q)}_{\lambda}$, show
the same behavior as we find in case of tightly binding potentials: The 
classical period $T_{\lambda}^{(cl)}$ increases whereas, the 
quantum revival time $T_{\lambda}^{(Q)}$ decreases as 
the external modulation strength $\lambda$ grows. 
However, for fixed values of ${\bar n}$ and $k^{\hspace{-2.1mm}-}$, 
the change in recurrence times is larger than what we find in the case
of tightly binding potentials. For the smaller values of $\rho$, we find that the classical period
and the quantum revival time in the presence and in the absence of external driving force
are related as
\begin{equation}
\frac{T_{\lambda}^{(cl)}}{T_0^{(cl)}}-\frac{T_{\lambda}^{(Q)}}{4T_0^{(Q)}}=0.
\end{equation}

A special situation occur at $\rho=2$ at which classical period and
the quantum revival time in the presence and in the absence of an
external modulation become infinite. From Eq.~(\ref{vkx}) we find
that in this case the potential energy is constant and independent of 
position. This corresponds to the
case of a free particle moving with a kinetic energy and a constant
potential energy.

In contrast to the tightly binding potentials, weakly binding
potentials display a direct proportionality between the mean 
quantum number $\bar n$ and 
the classical period $T_0^{(cl)}$ in the absence of
modulation, such that, $T_0^{(cl)}\propto \left(\bar
n +\gamma/4\right)^{\frac{\rho}{4-\rho}}$. For the reason, they
display a larger classical period and consequently a smaller
frequency as $\bar{n}$ increases. We find this dependence, for
example, in the Fermi accelerator~\cite{SaifR}.

The quantum revival time in the absence of external modulation,
$T_0^{(Q)}$, is directly proportional to the mean quantum number,
$\bar{n}$, in case of loosely binding potentials, such that,
$T_0^{(Q)}\propto \left(\bar n
+\gamma/4\right)^{\frac{4}{4-\rho}}$. The proportionality of
$T_0^{(cl)}$ and $T_0^{(Q)}$ on the mean quantum number $\bar n$
indicates that the change of $T_0^{(Q)}$, is greater as compared
with $T_0^{(cl)}$, as the value of $\bar n$ varies.
Interestingly the ratio between $T_0^{(Q)}$ and $T_0^{(cl)}$ remains the 
same both in the tightly binding potentials and loosely binding potentials,
however, it is not the same for the times in the presence of an 
external modulation.

For fixed values of $\rho$, and $\bar{n}$, in contrast to
the tightly binding potentials, we note that $T^{(cl)}_{0}$, and
$T^{(Q)}_{0}$, are directly proportional to the effective Planck's
constant, $k^{\hspace{-2.1mm}-}$. Therefore, for a fixed value of
$\bar n$ and $\rho$ the recurrence times increase with the scaled
Planck's constant, such that,
${k^{\hspace{-2.1mm}-}}^{\rho/(4-\rho)}$.

%\section{Modulated Gravitational Cavity}
%
%\label{sec:mgc}
%
As an example of loosely binding potentials, we study the linear potential
i.e. $k=1$. The energy eigen value of the linear gravitational potential 
truncated at the origon is given as
\begin{equation}
E_{n}^{(1)}=\left[\left(n+\frac{\gamma }{4}\right)\frac{3k^{\hspace{-2.1mm}-}\pi }{4\sqrt{2}}%
V_{0}\right]^{2/3}.  \label{Enk1}
\end{equation}%
The corresponding frequency $\omega $ is given by
\begin{equation}
\omega =\frac{2}{3{k^{\hspace{-2.1mm}-}}}\left(\bar{n}+\frac{\gamma }{4}%
\right)^{-1}E_{\bar{n}}^{(1)},  \label{omega1}
\end{equation}%
and nonlinearity $\zeta $ reads as%
\begin{equation}
\zeta =-\frac{2}{9{{k^{\hspace{-2.1mm}-}}^{2}}}\left(\bar{n}+\frac{\gamma
}{4}\right)^{-2}E_{\bar{n}}^{(1)}.  \label{zeta1}
\end{equation}

For linear gravitational potential, in the absence of external modulating
force, the classical period is given as
\begin{equation}
T_{0}^{(cl)}=\frac{3\pi k^{\hspace{-2.1mm}-}}{E_{\bar{n}}^{(1)}}\left(\bar{n}+%
\frac{\gamma }{4}\right),  \label{T0clk1}
\end{equation}%
and quantum recurrence time is given as
\begin{equation}
T_{0}^{(Q)}=\frac{18\pi k^{\hspace{-2.1mm}-}}{E_{\bar{n}}^{(1)}}\left(\bar{n}+%
\frac{\gamma }{4}\right)^{2}.  \label{T0quk1}
\end{equation}%
It is obvious from equations~(\ref{T0clk1}) and (\ref{T0quk1}) that
classical period and quantum recurrence time are directly proportional to the
mean quantum number $\bar{n}$ and varies as ${\bar n}^{1/3}$ 
and ${\bar n}^{4/3}$, respectively. 
%For unmodulated gravitational potential, the
%classical period and quantum recurrence times are calculated numerically in~%
%\cite{twen6}.

In the presence of external modulating force, the classical period $%
T_{\lambda }^{(cl)}$ is
\begin{equation}
T_{\lambda }^{(cl)}=\left\{ 1+\frac{1}{2}\left( \frac{\lambda V\Delta ^{2}}{%
2E_{\bar{n}}^{(1)}}\right) ^{2}\frac{1}{(1-\mu ^{2})^{2}}\right\}
T_{0}^{(cl)}\Delta ,  \label{Tclk1}
\end{equation}%
and quantum revival time reads as
\begin{equation}
T_{\lambda }^{(Q)}=\left\{ 1-\frac{1}{2}\left( \frac{\lambda V\Delta ^{2}}{%
2E_{\bar{n}}^{(1)}}\right) ^{2}\frac{3+\mu ^{2}}{(1-\mu ^{2})^{3}}\right\}
T_{0}^{(Q)},  \label{Tquk1}
\end{equation}%
where, the value of $\mu $ for gravitational potential reduces to $\mu={%
-N\Delta }/{6(\bar{n}+\frac{\gamma }{4})}$.

In the presence of external modulation, the classical period and the quantum
revival time have been studied for linear 
potential, which constitutes Fermi accelerator, 
numerically in the reference~\cite{twen3,twen4}.
The numerical results obtained in the earlier work show a very good agreement
with the analytical results presented in this paper. The analytical results 
lead to a better understanding of recurrence tracking microscope 
(RTM)~\cite{RTM} as well, which is to scan surface structures 
with nano-meter resolution.

\section{Acknowledgement}

SI wants to thank M. Ali, M. Ayub, R. ul Islam, and K.
Naseer for many discussions. He also submit his thanks to HEC
for funding through grant 041203721E-026. FS thanks 
G. Alber, and S. Watanabe for many useful discussions. He also thanks
Higher Education Commission, Pakistan (research grant R\&D/03/143), 
Quaid-i-Azam University and Fulbright Foundation, USA, for partial financial 
assistance.
FS thanks Prof. Dr. P. Meystre for his hospitality at the department of Physics,
University of Arizona where a part of the work was finished.

\end{document}